\documentclass[aps,prl,superscriptaddress,floatfix,preprint]{revtex4}
\usepackage[dvipdfmx]{graphicx}
\usepackage[dvipdfmx]{color}
\usepackage{amsmath}
\usepackage{amssymb}
\usepackage{dcolumn}
\usepackage{bm}
\usepackage{latexsym} 
\usepackage{setspace}
\usepackage{graphicx}
\usepackage{xcolor}
\usepackage{ulem}

\begin{document}
\title{Spin Seebeck effect in the layered ferromagnetic insulators CrSiTe$_3$ and CrGeTe$_3$}
\author{Naohiro Ito}
\email{nao.ito@cmpt.phys.tohoku.ac.jp}
\affiliation{Institute for Materials Research, Tohoku University, Sendai 980-8577, Japan}
\author{Takashi Kikkawa} 
\affiliation{Institute for Materials Research, Tohoku University, Sendai 980-8577, Japan}
\affiliation{WPI Advanced Institute for Materials Research, Tohoku University, Sendai 980-8577, Japan}
\author{Joseph Barker}
\affiliation{Institute for Materials Research, Tohoku University, Sendai 980-8577, Japan}
\affiliation{School of Physics and Astronomy, University of Leeds, Leeds LS2 9JT, United Kingdom}
\author{Daichi Hirobe} 
\email[Present address: Institute for Molecular Science, Okazaki, Aichi 444-8585, Japan.]{}
\affiliation{Institute for Materials Research, Tohoku University, Sendai 980-8577, Japan} 
\author{Yuki Shiomi}
\affiliation{Department of Basic Science, The University of Tokyo, Tokyo 153-8902, Japan}
\author{Eiji Saitoh}
\affiliation{Institute for Materials Research, Tohoku University, Sendai 980-8577, Japan}
\affiliation{WPI Advanced Institute for Materials Research, Tohoku University, Sendai 980-8577, Japan}
\affiliation{Department of Applied Physics, The University of Tokyo, Tokyo 113-8656, Japan}
\affiliation{Center for Spintronics Research Network, Tohoku University, Sendai 980-8577, Japan}
\affiliation{Advanced Science Research Center, Japan Atomic Energy Agency, Tokai 319-1195, Japan}
\date{\today}
\begin{abstract}
We have studied the longitudinal spin Seebeck effect (LSSE) in the layered ferromagnetic insulators CrSiTe$_3$ and CrGeTe$_3$ covered by Pt films in the measurement configuration where spin current traverses the ferromagnetic Cr layers. The LSSE response is clearly observed in the ferromagnetic phase and, in contrast to a standard LSSE magnet Y$_3$Fe$_5$O$_{12}$, persists above the critical temperatures in both CrSiTe$_3$/Pt and CrGeTe$_3$/Pt samples. With the help of a numerical calculation, we attribute the LSSE signals observed in the paramagnetic regime to exchange-dominated interlayer transport of in-plane paramagnetic moments reinforced by short-range ferromagnetic correlations and strong Zeeman effects. 
\end{abstract}
\maketitle
%
The longitudinal spin Seebeck effect (LSSE) generates spin currents in magnetic materials when a temperature gradient is applied \cite{LSSE}. By injecting this spin current into a paramagnetic metal it can be measured as a voltage through the inverse spin-Hall effect. Because of the simple bilayer structure needed to generate a thermoelectric voltage, LSSE devices have a potential use as thermoelectric conversion elements \cite{SSE-Appli-1,SSE-Appli-2}. From the point of basic physics, the LSSE is sensitive to spin correlations \cite{Xiao2010PRB, spin-correlation}, and thus can be exploited as a probe to study the dynamical spin susceptibility in magnetic materials \cite{adachi, Geprags2016NatCommun}. The LSSE was originally found in ferro(ferri)magnets and later measured also in antiferromagnets \cite{Antiferro-SSE-1,Antiferro-SSE-2,Antiferro-SSE-3,Antiferro-SSE-4} and paramagnets \cite{Para-SSE-1,Para-SSE-2}. However, the LSSE has not been studied in magnetic materials with two-dimensional (2D) crystal structures, even though 2D materials, such as transition-metal chalcogenides, have drawn extensive research attention due to their extraordinary magnetic properties \cite{2D-review}. \par
The layered ferromagnetic insulators CrSiTe$_3$ and CrGeTe$_3$ have been studied recently due to their intriguing physical properties. First-principles calculations \cite{First-Principle-1,First-Principle-2} predicted that the ferromagnetism in CrSiTe$_3$ survives even down to a monolayer thickness and indeed ferromagnetism in bilayer flakes of CrGeTe$_3$ was experimentally confirmed \cite{Two-layer-Ferro}. The Cr layers possess a graphenelike honeycomb structure and exotic properties are expected to arise in heterostructures fabricated with related 2D materials by van der Waals epitaxy. Furthermore, it has been reported that CrGeTe$_3$ acts as an ideal ferromagnetic substrate for the growth of the popular topological insulator Bi$_2$Te$_3$ \cite{Hetero}. \par
\begin{figure}[t]
\begin{center}
\includegraphics[width=8.5cm]{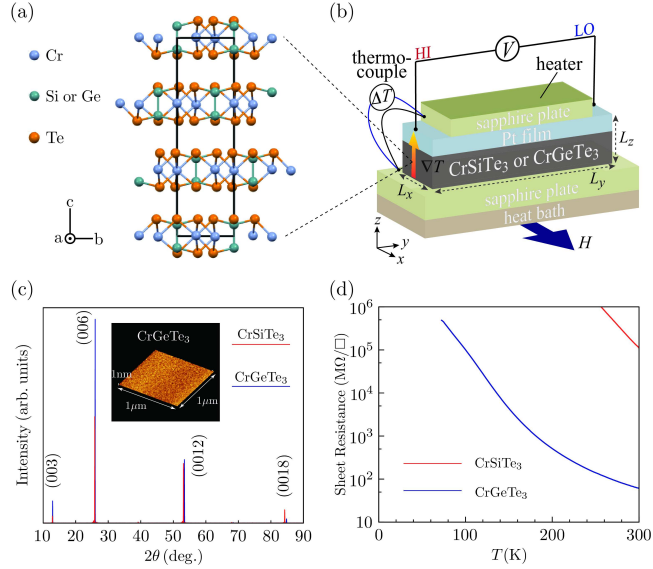}
\end{center}
\caption{(a) Schematic illustration of the crystal structure of CrSiTe$_3$ and CrGeTe$_3$.  (b) Schematic illustration of the LSSE measurements. $H$ denotes the external magnetic field and $\Delta T$ ($\nabla T$) the temperature difference (gradient). (c) X-ray diffraction patterns of CrSiTe$_3$ and CrGeTe$_3$ single crystals. The inset is an atomic force microscope image of the surface of CrGeTe$_3$. (d) $T$ dependence of the sheet resistance for CrSiTe$_3$ and CrGeTe$_3$. 
}
\label{fig:fig1}
\end{figure}

In this Rapid Communication, we studied the LSSE \cite{LSSE} in the ferromagnets CrSiTe$_3$ and CrGeTe$_3$ in contact with Pt films. The crystal structure of CrSiTe$_3$ and CrGeTe$_3$ is illustrated schematically in Fig. \ref{fig:fig1}(a). The Cr$^{3+}$ (spin 3/2) ions form a honeycomb lattice in the $ab$ plane with the Si or Ge atoms in the center of the hexagon and the Cr$^{3+}$ atoms are surrounded by octahedra of Te atoms \cite{CST-struc-1,CST-struc-2,CGT-struc-TC}. The honeycomb layers stack along the $c$ axis, held together by van der Waals interactions, forming a quasi-2D structure with a highly anisotropic magnetic environment. Interestingly, it was reported that CrSiTe$_3$, with a Curie temperature of $T_{\rm C} \approx 31$ K, has short-range, in-plane ferromagnetic correlations which survive up to at least 300 K, whereas out-of-plane correlations disappear above 50 K \cite{Neutron-Scatter-2}. This is because the in-plane exchange coupling is more than five times greater than the out-of-plane coupling \cite{Neutron-Scatter-2}. \par
Single crystals of CrSiTe$_3$ and CrGeTe$_3$ were grown by a self-flux method, following the procedure described in the literature \cite{Hetero,CST-CrystalGrowth-TC,CGT-CrystalGrowth-TC}. First, high-purity powders of Cr, Si, Ge, and Te were placed in alumina crucibles in a molar ratio of Cr:Si:Te = 1:2:6 and Cr:Ge:Te = 1:3:18; the excess Si, Ge, and Te work as a flux for the crystal growth. The alumina crucibles were placed inside quartz tubes and sealed under argon atmosphere (pressure of 0.3 bar). The ampoules were then heated up to 1150 $^\circ$C (700 $^\circ$C), and maintained at theses temperatures for 16 h (22 days) and then slowly cooled to 700 $^\circ$C (500 $^\circ$C) for CrSiTe$_3$ (CrGeTe$_3$), followed by centrifugation to remove excess flux. Single crystals were obtained as platelike forms with the size of several millimeters. \par
The sample structures were characterized by x-ray diffraction with Cu $K\alpha$1 radiation at room temperature. Figure \ref{fig:fig1}(c) shows x-ray diffraction patterns of the CrSiTe$_3$ and CrGeTe$_3$ samples. Both samples show only sharp $(00n)$ peaks; no impurity peaks were observed. The widest planes of the single crystals were determined as crystallographic $ab$ planes. The lattice parameter was estimated to be $c$ = 20.67 $\rm\AA$ for CrSiTe$_3$ and $c$ = 20.56 $\rm\AA$ for CrGeTe$_3$, consistent with previous reports \cite{CST-struc-1,CGT-struc-TC,CST-CrystalGrowth-TC,Neutron-Scatter,CGT-Mag}. \par
The temperature ($T$) dependences of the in-plane sheet resistance of CrSiTe$_3$ and CrGeTe$_3$ crystals are shown in Fig. \ref{fig:fig1}(d). For CrSiTe$_3$, the sheet resistance increases with decreasing $T$, and goes beyond the measurement limit around $250$ K; at lower temperatures, the CrSiTe$_3$ samples can be viewed as a good insulator. The sheet resistance of CrGeTe$_3$ shows a similar $T$ dependence, but is three orders of magnitude smaller than that of CrSiTe$_3$. In CrGeTe$_3$/Pt devices ferromagnetic itinerant transport, such as the anomalous Nernst effect \cite{ANE}, is expected to be negligible below $\approx$ 100 K. \par
\begin{figure}[t]
\begin{center}
\includegraphics[width=8.5cm]{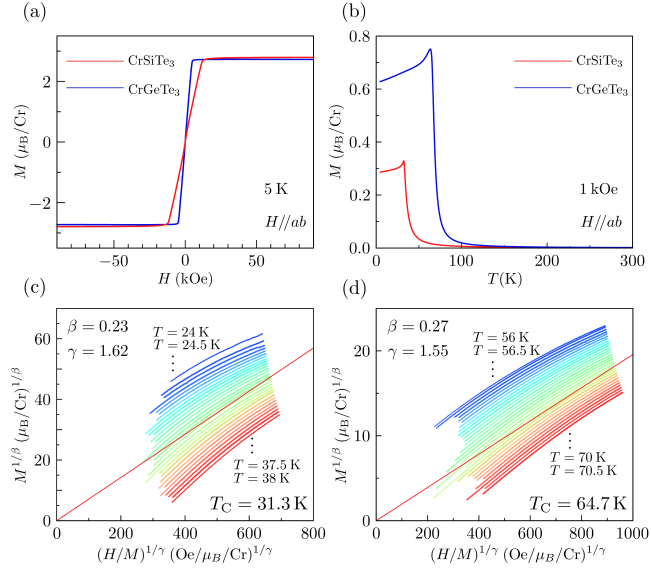}
\end{center}
\caption{
(a) $H$ dependence of $M$ ($M$-$H$ curve) for CrSiTe$_3$ and CrGeTe$_3$ at 5 K. (b) $T$ dependence of $M$ for CrSiTe$_3$ and CrGeTe$_3$ at 1 kOe (below saturation). (c),(d) Modified Arrott plot of isotherms for (c) CrSiTe$_3$ and (d) CrGeTe$_3$, from which the $T_{\textrm C}$ values are determined. The red straight lines in (c) and (d) are Eq.(\ref{eq:eq1}) at $T=T_{\rm C}$.
}
\label{fig:fig2}
\end{figure}

 
Magnetic properties were measured with a vibrating sample magnetometer. Figure \ref{fig:fig2}(a) shows the magnetization $M$ versus magnetic field $H$ applied in the $ab$ plane of CrSiTe$_3$ and CrGeTe$_3$ crystals at $T = 5~\textrm{K}$.
A magnetization curve with a saturation magnetization of about 2.7$\mu_{\rm B}$ was observed for both samples, in good agreement with the spin 3/2 of Cr$^{3+}$ ions. In Fig. \ref{fig:fig2}(b), we show the $T$ dependence of $M$ for CrSiTe$_3$ and CrGeTe$_3$ at $H = 1~\textrm{kOe}$. A sharp paramagnetic to ferromagnetic phase transition was observed in both samples \cite{CST-CrystalGrowth-TC,CGT-Mag}. \par
We evaluated the critical temperature $T_{\textrm C}$ for CrSiTe$_3$ and CrGeTe$_3$ using a modified Arrott plot analysis \cite{CGT-Mag,Magnetization-Measure,CST-Mag}. This method determines the critical exponents $\beta$ and $\gamma$ by the fitting of $H/M$ and $M$ with the Arrott-Noaks equation:
\begin{align}
\left(\frac{H}{M}\right)^{1/\gamma} = a\frac{T-T_{\rm C}}{T} + bM^{1/\beta},
\label{eq:eq1}
\end{align}
where $a$ and $b$ are fitting parameters.  
Fitting of Eq. (\ref{eq:eq1}) with a rigorous iterative method \cite{CGT-Mag,Magnetization-Measure,CST-Mag} gives $T_{\rm C} = 31.3~\textrm{K}$ for CrSiTe$_3$ and $T_{\rm C} = 64.7~\textrm{K}$ for CrGeTe$_3$ [see Figs. \ref{fig:fig2}(c) and \ref{fig:fig2}(d), where the estimated critical exponents $\beta$ and $\gamma$ are also shown].
The values of $\beta$ are smaller than in three-dimensional (3D) spin models ($\beta_{{\rm 3D\:Heisenberg}}=0.365$ and $\beta_{{\rm 3D\:Ising}}=0.325$) but larger than 2D spin models ($\beta_{{\rm 2D\:Ising}}=0.125$), signaling a 2D character of the ferromagnetic transition \cite{CGT-Mag,CST-Mag}.\par
To measure the LSSE we used samples of CrSiTe$_3$ (CrGeTe$_3$) with dimensions $L_x = 1.5~\textrm{mm}$ $(0.8~\textrm{mm})$, $L_y = 3.5~\textrm{mm}$ $(4.4~\textrm{mm})$, and $L_z = 0.2~\textrm{mm}$ $(75~\mu \textrm{m})$ and deposited a 5-nm-thick Pt film on the surface [see Fig. \ref{fig:fig1}(b)]. To ensure clean and flat interfaces, their (as-grown) top ($ab$-plane) surfaces were exfoliated using adhesive tape before the Pt deposition; the resultant surface roughnesses ($R_a$) of the CrSiTe$_3$ and CrGeTe$_3$ samples were $5.4 \times 10^{-2}~\textrm{nm}$ and $4.1 \times 10^{-2}~\textrm{nm}$, respectively, confirming the samples are very flat and smooth [see the atomic force microscope image for the CrGeTe$_3$ surface shown in the inset to Fig. \ref{fig:fig1}(c)]. To apply a temperature gradient, $\nabla T$, along the $c$ axis [$z$ axis in  Fig. \ref{fig:fig1}(b)] of the samples, the CrSiTe$_3$/Pt and CrGeTe$_3$/Pt were sandwiched between two sapphire plates; a $100~\Omega$ chip resistor was fixed on one plate, while the other is connected to a heat bath [see Fig. \ref{fig:fig1}(b)]. By applying a charge current to the resistor, a constant temperature difference, $\Delta T$, of $2~\textrm{K}$ ($1~\textrm{K}$) was generated for the CrSiTe$_3$/Pt (CrGeTe$_3$/Pt) sample that was measured by using type-E thermocouples attached to the two sapphire plates [see Fig. \ref{fig:fig1}(b)]. The external magnetic field $H$ was applied in the $ab$ plane (along the $x$ axis) and the thermal voltage $V$ between the ends of the Pt film (along the $y$ axis; its distance $L_y$) was measured. We define the LSSE voltage $V_{\rm LSSE}$ as the antisymmetric contribution of the thermoelectric voltage: $[V(+H)-V(-H)]/2$. Hereafter, we mainly use the transverse thermopower $S = (V_{\textrm{LSSE}}/\Delta T)(L_z/L_y)$ as the normalized LSSE voltage. \par
\begin{figure}[t]
\begin{center}
\includegraphics[width=8.5cm]{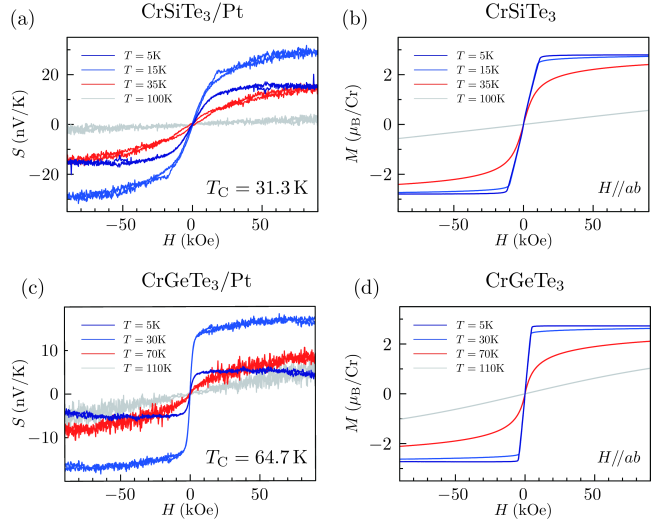}
\end{center}
\caption{
$H$ dependence of the normalized LSSE signal $S$ in (a) CrSiTe$_3$/Pt and (c) CrGeTe$_3$/Pt and $H$ dependence of $M$ for (b) CrSiTe$_3$ and (d) CrGeTe$_3$ at several temperatures, where $H$ was applied in the $ab$ plane and swept between $\pm$90 kOe. 
}
\label{fig:fig3}
\end{figure}

Figures \ref{fig:fig3}(a) and \ref{fig:fig3}(c) show the $H$ dependence of $S$ in the CrSiTe$_3$/Pt and CrGeTe$_3$/Pt samples at selected temperatures. 
In the ferromagnetic phase below $T_{\rm C}$ clear LSSE signals were observed for both the samples [see the dark- and bright-blue solid lines in Figs. \ref{fig:fig3}(a) and \ref{fig:fig3}(c)].
The $S$-$H$ curves for the CrSiTe$_3$/Pt and CrGeTe$_3$/Pt qualitatively agree with the $M$-$H$ curves for the bulk CrSiTe$_3$ and CrGeTe$_3$ crystals as shown in Figs. \ref{fig:fig3}(b) and \ref{fig:fig3}(d); by increasing $H$ from zero, the $S$ amplitude rapidly increases and almost saturates above the magnetization saturation field $H_{\rm c}$ for $M$ ($\sim 15~\textrm{kOe}$ for CrSiTe$_3$ and $\sim 6~\textrm{kOe}$ for CrGeTe$_3$). This is a characteristic feature of the LSSE \cite{SSE-Appli-2}. \par
Interestingly, clear LSSE signals are observed at the paramagnetic phases above $T_{\rm C}$ where long-range ferromagnetic ordering is absent.
The red solid lines in Figs. \ref{fig:fig3}(a) and \ref{fig:fig3}(c) represent the $S$-$H$ curves for the CrSiTe$_3$/Pt and CrGeTe$_3$/Pt samples at 35 and 70 K, higher than their $T_{\rm C}$ values of $31.3$ and $64.7~\textrm{K}$, respectively. We attribute the nonlinear $H$ dependence of $S$ to the LSSE as the observed $S$ signals should originate purely from the LSSE in the paramagnetic phase since the Nernst effect in CrSiTe$_3$ and CrGeTe$_3$ will be vanishingly small for the high resistivities at these temperatures.  \par
\begin{figure}[t]
\begin{center}
\includegraphics[width=8.5cm]{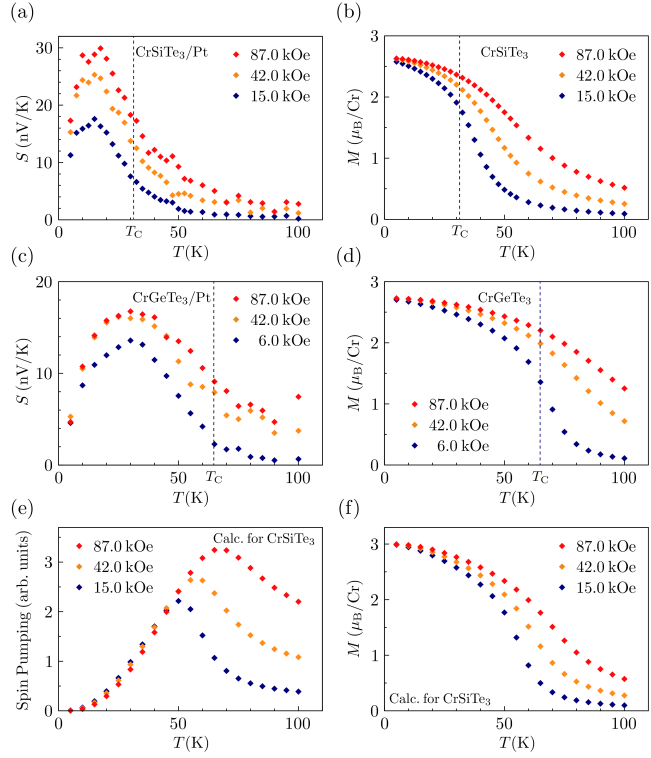}
\end{center}
\caption{
$T$ dependence of the normalized LSSE signal $S$ (a) at 15.0, 42.0, and 87.0 kOe in CrSiTe$_3$/Pt and (c) at 6.0, 42.0, and 87.0 kOe in CrGeTe$_3$/Pt.
$T$ dependence of the magnetization $M$ (b) at 15.0, 42.0, and 87.0 kOe in CrSiTe$_3$ and (d) at 6.0, 42.0, and 87.0 kOe in CrGeTe$_3$. 
(e) Calculated spin pumping and (f) magnetization for CrSiTe$_3$ at 15.0, 42.0, and 87.0 kOe using the Hamiltonian in Ref. \cite{Neutron-Scatter-2}. 
}
\label{fig:fig4}
\end{figure}

We systematically measured the $T$ dependence of the LSSE. The dark-blue plots in Figs. \ref{fig:fig4}(a) and \ref{fig:fig4}(c) show the $S$ versus $T$ results for the CrSiTe$_3$/Pt and CrGeTe$_3$/Pt samples, at a low field of $H = 15$ and $6~\textrm{kOe}$, respectively. By increasing $T$ from low temperature, $S$ increases and takes a maximum value at around 15 K (30 K) for the CrSiTe$_3$/Pt (CrGeTe$_3$/Pt) sample. Further increasing $T$, $S$ decreases. The behavior is similar to the LSSE in a 3D ferro(ferri)magnet Y$_3$Fe$_5$O$_{12}$ (YIG) \cite{peak-structure,Jin2015PRB,High-Temp-SSE}. 
In contrast to the YIG case, however, the $S$ signal persists at and above $T_{\rm C}$ [see the dark-blue plots around dashed lines in Figs. \ref{fig:fig4}(a) and \ref{fig:fig4}(c)]. Upon further increasing $T$, the LSSE signal disappears around 50 K (90 K) for the CrSiTe$_3$/Pt (CrGeTe$_3$/Pt) sample, which is higher than $T_{\rm C}$ by $\sim 20~\textrm{K}$.\par
\begin{figure}[t]
\begin{center}
\includegraphics[width=8.5cm]{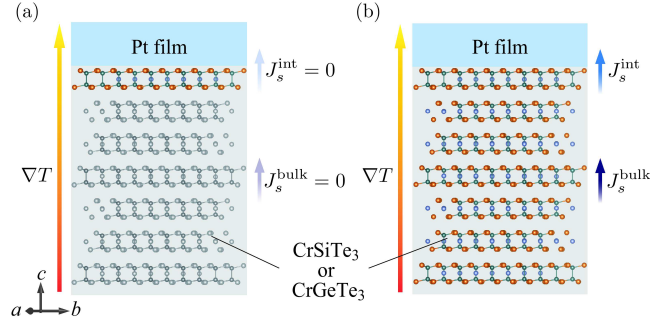}
\end{center}
\caption{
Schematic illustrations of CrSiTe$_3$/Pt or CrGeTe$_3$/Pt under $\nabla T$.
(a) For $T \gg T_{\rm C}$, interfacial spin pumping from the in-plane spin correlations in CrSiTe$_3$ and CrGeTe$_3$ adjacent to the Pt interface is absent ($J_{s}^{\rm int} = 0$), where bulk spin transport ($\parallel \nabla T$) in CrSiTe$_3$ and CrGeTe$_3$ is inert ($J_{s}^{\rm bulk} = 0$) since thermal fluctuations mask weak out-of-plane spin-exchange interaction $J_{\rm c}$ at such high temperatures. (b) For $T \leqq T_{\rm C}$ and the paramagnetic phase just above $T_{\rm C}$, bulk spin transport between the planes becomes active ($J_{s}^{\rm bulk} \neq 0$) and creates nonequilibrium magnon population at the interface, causing finite interfacial spin pumping $J_{s}^{\rm int}$. 
}
\label{fig:fig5}
\end{figure}

The observation of the LSSE in the paramagnetic phase above $T_{\rm C}$ gives an insight into the different roles of the anisotropic (in-plane and out-of-plane) spin correlations in LSSE. 
The quasi-2D ferromagnetism of CrSiTe$_3$ and CrGeTe$_3$ is due to the significant difference in strength of the in-plane and out-of-plane exchange interactions \cite{Two-layer-Ferro,Neutron-Scatter-2}. In CrSiTe$_3$ the in-plane exchange coupling strength $J_{ab} \sim 15~\textrm{K}$ is more than five times larger than the out-of-plane $J_{c}$ \cite{Neutron-Scatter-2}. Short-range in-plane ferromagnetic correlations even persist to room temperature, while the out-of-plane correlations rapidly diminish above $T_{\rm C}$ \cite{Neutron-Scatter-2}. 
If the LSSE we measured was driven by spin pumping from the in-plane ($\perp \nabla T$) spin correlations adjacent to the Pt interface, the LSSE signal would appear until 300 K. That is not the case [Fig. \ref{fig:fig5}(a)]. Our experimental results show the vital role of spin transport between the planes (along the temperature gradient, $\parallel \nabla T$) to create the nonequilibrium magnon population essential for the appearance of LSSE [Fig. \ref{fig:fig5}(b)] \cite{SSE-theory,Cornelissen2016PRB-chemical-potential}. This experimentally decouples the physics of interface spin pumping from the bulk spin transport [compare Figs. \ref{fig:fig5}(a) and \ref{fig:fig5}(b)]. Here, the interface spin pumping refers to the injection of spin currents by magnetization dynamics at the interface, and should thus be distinct from the bulk spin transport. This decoupling has been shown on picosecond timescales in extreme nonequilibrium using terahertz pulses \cite{Seifert2018NatCommun}. But we demonstrate this in a conventional LSSE experiment in a nonequilibrium but steady state.
\par
In the ferromagnetic phase below $T_{\rm C}$, the LSSE can be understood in the same manner as conventional 3D ferromagnets, such as YIG \cite{SSE-theory,Cornelissen2016PRB-chemical-potential}; in the long-range ordered state, the out-of-plane exchange coupling \cite{Two-layer-Ferro,Neutron-Scatter-2} facilitates the spin transport along the $c$ axis ($\parallel \nabla T$) [Fig. \ref{fig:fig5}(b)].
In the paramagnetic phase above $T_{\rm C}$, there is no magnetization (in zero magnetic field) and magnons cease to exist. Spin-current transport is therefore not usually observed \cite{shiomi-prl}. In the case of CrSiTe$_3$ and CrGeTe$_3$, however, strong short-range ferromagnetic correlations \cite{Two-layer-Ferro,Neutron-Scatter-2} above $T_{\rm C}$ form in the $ab$ plane. These spin correlations can be conveyed along the $c$ direction [Fig. \ref{fig:fig5}(b)] via spin-exchange coupling ($J_{c}$ \cite{Two-layer-Ferro,Neutron-Scatter-2}) \cite{Bennett1965PhysRev,Wesenberg2017NatPhys} and dipole-dipole interactions \cite{Oyanagi}, giving a finite inverse spin-Hall voltage when a moderate $H$ is applied to align spins. Since the LSSE response is observed in a limited $T$ range above $T_{\rm C}$ [Figs. \ref{fig:fig4}(a) and \ref{fig:fig4}(c)], the exchange coupling is likely to be the primary interaction mediating the interlayer spin transport. 
\par 
We also examined the $T$ dependence of the LSSE response with high magnetic fields up to $87.0~\textrm{kOe}$ in Figs. \ref{fig:fig4}(a) and \ref{fig:fig4}(c) (see orange and red dots). As $H$ increases $S$ also increases in magnitude and survives to higher temperatures. For CrSiTe$_3$/Pt [Fig. \ref{fig:fig4}(a)] the $S$ signal at $87.0~\textrm{kOe}$ appears even at $\sim 80~\textrm{K}$, more than twice the critical temperature $T_{\rm C} \sim 31.3~\textrm{K}$ \cite{comment}. Under strong magnetic fields, the spin polarization is further enhanced by the Zeeman interaction, which may be responsible for the $S$ increase. This breaks the symmetry of the Hamiltonian so the system no longer exhibits true critical behavior \cite{Goldenfeld}. This is seen in the $M$-$T$ curves shown in Figs. \ref{fig:fig4}(b) and \ref{fig:fig4}(d) where the ferroparamagnetic transition becomes less defined and there is a significant magnetization above the true critical temperature. $S$ decreases more rapidly than $M$ with increasing $T$ [Figs. \ref{fig:fig4}(a) and \ref{fig:fig4}(c)]. The spin Seebeck signal has almost disappeared even though $M > 0.5\mu_{\mathrm{B}}$, which also points to the importance of the out-of-plane spin transport in the LSSE. The coupling between the layers ($J_{c}$) is insensitive to magnetic fields so the decay rate of $S$ is not relational as with applied fields. Hence $S$ does not show the dependence on $M$ that would be expected for a ``3D" magnetic system.
\par
To confirm the above scenario and separate the spin pumping and spin transport of the LSSE we performed atomistic spin dynamics calculations of the (in-plane) magnetization and spin pumping (in the absence of spin transport) \cite{footnote-comment}. This provides the pure spin pumping contribution of the LSSE without concerns about the magnon distribution at the interface [its formulation is given by Eq. (6) in Ref. \cite{spin-correlation}]. The Hamiltonian is based on the magnetic parameters of CrSiTe$_{3}$ reported in Ref. \cite{Neutron-Scatter-2}. Our calculations show a good agreement with the $M$-$T$ curves of the experiments [Figs. \ref{fig:fig4}(b) and \ref{fig:fig4}(d)], although with a slightly higher critical temperature. Figure \ref{fig:fig4}(e) shows $T$ dependence of the spin pumping amplitude at $15$, $42$, and $87$ kOe. Large $H$ increases the spin pumping significantly at temperatures far above $T_{\rm C}$ due to the Zeeman effect; it shows a maximum at approximately 70 K and decreases gently. At 100~K there is still a large difference between the different field values, roughly proportional to $M$. This is also consistent with analytic approaches in paramagnets \cite{okamoto}. The reason this is not seen in our experimental measurements is because the out-of-plane spin transport between the layers has vanished and so the magnon distribution at the interface is close to equilibrium. Therefore the paramagnetic spin pumping would not be observed.
\par 
To summarize, we studied the LSSE in the ferromagnetic transition-metal trichalcogenides CrSiTe$_3$ and CrGeTe$_3$ with Pt contact. In contrast to typical LSSE magnet YIG, the LSSE signal in the CrSiTe$_3$/Pt and CrGeTe$_3$/Pt was found to persist even above $T_{\rm C}$. The LSSE above $T_{\rm C}$ is dominated by thermal spin transport via out-of-plane exchange coupling between the 2D layers. At low magnetic fields the strong short-range ferromagnetic correlations inherent in CrSiTe$_3$ and CrGeTe$_3$ are the main spin-current carriers, while the Zeeman energy, comparable in strength to the exchange coupling, is dominant for the spin alignment under high magnetic fields. Our numerical simulation corroborates the strong magnetic-field effects and points out the importance of the interlayer spin transport regardless of the quasi-2D structure. The measurable spin transport across the ferromagnetic Cr layers in CrSiTe$_3$ and CrGeTe$_3$ could be useful in studying spin transports in heterostructures fabricated with other 2D materials and topological insulators.
\par
The authors thank K. Ohgushi, K. Emi, K. Tanigaki, T. Ogasawara, K. Nawa, and M. Takahashi for their valuable comments on crystal growth.
The authors also thank Y. Nambu, S. Takahashi, and K. Oyanagi for their fruitful discussion.
This work was supported by JST ERATO ``Spin Quantum Rectification Project'' (Grant No. JPMJER1402), JSPS KAKENHI (Grants No. JP26103005, No. JP19H02424, No. JP19K21039, No. JP18H04215, No. JP18H04311, No. JP19K21031, and No. JP17K14102), JSPS Core-to-Core program ``the International Research Center for New-Concept Spintronics Devices'', and GP-Spin, Tohoku University. J.B. acknowledges support from the Royal Society through a University Research Fellowship. D.H. was supported by the Yoshida Scholarship Foundation through the Doctor 21 program.

\newpage

\begin{thebibliography}{99}
\bibitem{comment} $S$ curves become linear with respect to $H$ above 65 K for the CrSiTe$_3$/Pt and 85 K for CrGeTe$_3$/Pt  [see the $S(H)$ result at 100 K shown in Figs. \ref{fig:fig3}(a) and \ref{fig:fig3}(c)]. Above these temperatures, we cannot distinguish the contribution from an $H$-linear normal Nernst effect to the $S$ signal \cite{ANE}. 
We thus avoid a detailed discussion on the high-$H$-induced $S$ enhancement above 65 K for CrSiTe$_3$/Pt and above 85 K for CrGeTe$_3$/Pt.
\end{thebibliography}
\end{document}